\documentclass[11pt]{article}

 \usepackage{amssymb}
\usepackage{amsthm}
\usepackage{epsfig}
\usepackage{color}
\usepackage{fullpage}
\usepackage{algorithm}
\usepackage[noend]{algorithmic}
\usepackage{appendix}
\usepackage{times}

\newtheorem{theorem}{Theorem}

\newtheorem{lemma}[theorem]{Lemma}

\newtheorem{claim}[theorem]{Claim}

\newtheorem*{uc}{Uniqueness Claim}

\title{The Complexity of Guarding Terrains}

\author{\begin{tabular}[t]{c@{\extracolsep{8em}}c}
           James King  & Erik Krohn\\
\it        McGill University & \it University of Iowa\\
\it        jking@cs.mcgill.ca & \it erik-krohn@uiowa.edu\\
\end{tabular}}

\begin{document}

\maketitle


\begin{abstract}
A set $G$ of points on a 1.5-dimensional terrain, also known as an $x$-monotone polygonal chain, is said to guard the terrain if any point on the terrain is ‘seen’ by a point in $G$. Two points on the terrain see each other if and only if the line segment between them is never strictly below the terrain. The minimum terrain guarding problem asks for a minimum guarding set for the given input terrain. We prove that the decision version of this problem is NP-hard. This solves a significant open problem and complements recent positive approximability results for the optimization problem.

Our proof uses a reduction from PLANAR 3-SAT. We build gadgets capable of ‘mirroring’ a consistent variable assignment back and forth across a main valley. The structural simplicity of 1.5-dimensional terrains makes it difficult to build general clause gadgets that do not destroy this assignment when they are evaluated. However, we exploit the structure in instances of PLANAR 3-SAT to find very specific operations involving only ‘adjacent’ variables. For these restricted operations we can construct gadgets that allow a full reduction to work.
\end{abstract}

\thispagestyle{empty}
\newpage
\setcounter{page}{1}

\section{Introduction}

An instance of the {\em terrain guarding problem} contains a terrain $T$ that is an $x$-monotone polygonal chain.  An $x$-monotone chain in $\Re^2$ is a chain that intersects any vertical line at most once.  The terrain is given by its set of vertices $P = \lbrace v_1, v_2, ..., v_n\rbrace$, where $v_i = \left(x_i, y_i\right)$.  The vertices are ordered such that $x_i < x_{i+1}$.  There is an edge connecting each $\left(v_i, v_{i+1}\right)$ pair where $i=1, 2, ..., n-1$.  We say a point $p$ on the terrain \textit{sees} another point $q$ on the terrain if the line segment $\overline{pq}$ is never strictly below the terrain $T$.

A set $G$ of points on the terrain is called a \emph{guarding set} if every point on the terrain is seen by some point in $G$.  The optimization version of the terrain guarding problem is the problem of finding a minimum guarding set for a given terrain.  There are two standard versions of the terrain guarding problem: a discrete version and a continuous version.  The discrete version allows us to place guards only at the vertices of the terrain.  The continuous version, which we have defined above, allows guards to be placed anywhere on the terrain.  In other versions a subset of points on the terrain to guard is given with the input.

Motivation for guarding terrains comes from scenarios that include covering a road with street lights or security cameras.  Other applications include finding a configuration for line-of-sight transmission networks for radio broadcasting, cellular telephony and other communication technologies \cite{MKM05}.

The complexity of terrain guarding has been an open problem of interest since 1995, when an NP-completeness proof was proposed but never completed by Chen \textit{et al.} \cite{CEU96}.  With the problem's hardness strongly suspected but not known, a series of approximation algorithms have been developed over the last decade.
The first constant factor approximation for the terrain guarding problem was shown by Ben-Moshe \textit{et al.} in \cite{MKM05}.  Clarkson and Varadarajan also give a constant factor approximation in \cite{CV05}.  A 4-approximation was proposed by King in \cite{K06} but further analysis increased the approximation factor to 5.  A 4-approximation was given by Elbassioni \textit{et al.} in \cite{E09}.  Recently a PTAS was given by Gibson {\em et al.} in \cite{G09}.  With the knowledge that the problem is not APX-complete, it is of even greater interest whether or not it is NP-complete, and this has been reiterated with each approximation algorithm developed.

The terrain guarding problem is closely related to the \emph{art gallery problem} that involves guarding the interior of a polygon.  The basic version of
the art gallery problem is that of {\em vertex guarding} a simple polygon, where we are given a
simple polygon and we wish to find the smallest subset of the vertices that see the entire polygon.
The {\em point guarding} version allows guards to be placed anywhere inside the polygon.

The art gallery problem was shown to be NP-complete by Lee and Lin in \cite{LL86}.  Along with being NP-complete, the art gallery problem was shown to be APX-hard in \cite{E98}. This means that there exists a constant $\epsilon > 0$ such that no polynomial time algorithm can guarantee an approximation ratio of $1+\epsilon$ unless $P=NP$.   Ghosh provides a $O(\log n)$-approximation for the problem of vertex guarding an $n$-vertex simple polygon \cite{G88}.  The point guarding problem seems to be much harder than the vertex guarding problem and precious little is known about it \cite{DKDS07}.  A restricted version of the point guarding problem where the polygon is $x$-monotone has been shown to have an $O(1)$-approximation by Nilsson in \cite{N05}.  Based on his result Nilsson also provides a $O(OPT^2)$ approximation for rectilinear polygons.

Straightforward attempts to show NP-hardness for the terrain guarding problem run up against the large amount of restriction in the complexity of terrains.  By far the most significant restriction is given by the following claim first noted by Ben-Moshe \textit{et al.} \cite{MKM05}:

\begin{claim}[Order Claim]
Let $a,b,c,d$ be four points on the terrain in increasing order of $x$-coordinates.  If $a$ sees $c$ and $b$ sees $d$, then $a$ sees $d$.  
\end{claim}

The order claim is crucially exploited by all approximation algorithms for the problem.  In this paper we develop a construction that overcomes the order claim obstacle and shows that the terrain guarding problem is NP-hard.  Therefore, an exact polynomial time algorithm is not possible unless ${\rm P=NP}$.  The NP-hardness result is shown for the standard discrete and continuous variants of the problem.  

According to Demaine and O'Rourke \cite{DO05}, the complexity of the terrain guarding problem was posed by Ben-Moshe. We quote from \cite{DO05}:
\begin{quote}
 What is the complexity of computing the guard set of minimum size for a given x-monotone chain in the plane? According to the poser, ``most tenured professors think the problem is NP-hard.'' This problem in fact goes back to 1995, when Chen \textit{et al.} \cite{CEU96} claimed an NP-hardness result, but ``the proof, whose details were omitted, was never completed successfully'' \cite{K06}.
\end{quote}

The remainder of the paper is organized as follows.  Our reduction is from planar 3SAT and is overviewed in Section 2.  Section 3 describes the gadgets used in the reduction and Section 4 provides a conclusion and future work.  The interested reader can see an example of a full reduction in Appendix C.

\section{Reduction: Overview}


The initial reduction will be for the discrete terrain guarding problem where both the set of guards and the set of points to be guarded are a finite subset of the terrain.  We give a reduction from the planar 3SAT problem.  This problem was shown to be NP-complete in \cite{L82}.  Planar 3SAT is defined as follows: Let $\Phi = (X,C)$ be an instance of 3SAT, with variable set $X=\{x_1,\ldots,x_n\}$ and clauses $C=\{c_1,\ldots,c_m\}$ such that each clause consists of exactly three distinct literals. Define a \textit{formula graph} $G_\Phi = (V,E)$ with vertex set $V = X\bigcup C$ and edges $E = E_1\bigcup E_2$ where $E_1 = \{(x_i,x_{i+1})|1\le i\le n\}$, and
$E_2 = \{(x_i,c_j) \ | \ c_j \mbox{ contains } x_i \mbox{ or } \overline{x_i}\}$. A 3SAT formula $\Phi$ is called \textit{planar} if the corresponding formula graph $G_\Phi$ is planar. The edge set $E_1$ defines a cycle on the vertices $X$, and thus divides the plane into exactly 2 faces. Each node $c_j\in C$ lies in exactly one of those two faces. We have to determine whether there exists an assignment of truth values to the variables in $X$ that satisfies all the clauses in the $C$.

It is easy to see that the clauses inside the variable cycle can be generated by performing a sequence $\beta$ of steps starting with $\sigma=\left< x_1,\ldots,x_n\right>$ where at each step we do one of the following until $\sigma$ becomes empty:

\begin{enumerate}
 \item Delete a variable from sequence $\sigma$ and call the resulting variable sequence $\sigma$.
 \item Generate a clause using three consecutive variables in $\sigma$ and delete the middle variable from $\sigma$.  Call the resulting variable sequence $\sigma$.
\end{enumerate}

Similarly there is a different sequence $\alpha$ of steps starting from $\sigma=\left<x_1,\ldots,x_n\right>$ that generates all clauses outside the variable cycle.  The interested reader can see Appendix C for an illustration of this sequence.

\begin{figure}[tpb]
\centering
\input{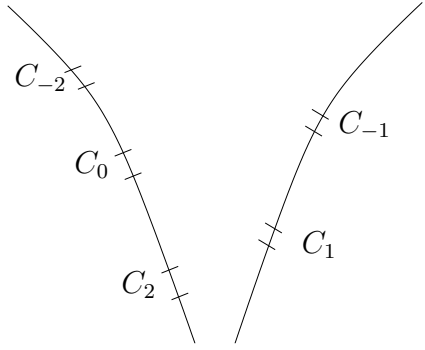}
\caption{A coarse view of a terrain $T$ constructed by our reduction showing chunks $C_{-2}, C_{-1}, C_0, C_1$ and $C_2$.}
\label{fig:overview}
\end{figure}

The terrain $T$ constructed by our reduction is shaped like a valley.  A coarse view of the terrain can be seen in Figure \ref{fig:overview}.  We identify disjoint pieces of the terrain called \textit{chunks}.  Even indexed chunks, $C_0, C_2, C_4,$ $..., C_{-2}, C_{-4}, ...$, are on the left side of the terrain and odd indexed chunks, $C_1, C_3, ..., C_{-1}, C_{-3}, ...$, are on the right side of the terrain.  Chunks $C_0, C_1, C_2, ..., C_k$ are used to ``implement'' the sequence $\beta$.  Chunks $C_0, C_{-1}, C_{-2}, ..., C_{-k'}$ are used to ``implement'' the sequence $\alpha$.

Recall that we are considering the discrete terrain guarding problem; we have a finite set of guards.  Chunks contain {\em distinguished points} which are the points to be guarded in our reduction.  Chunks also contain a set of potential guard locations.  Distinguished points and the set of potential guard locations will be defined in Section 3.


Corresponding to each even chunk, $C_i$ is a subsequence $\lambda_i$ of the sequence of variables $\langle x_1, x_2, ..., x_n \rangle$.  There will be $2|\lambda_i|$ guard locations, one for each of the $2|\lambda_i|$ literals corresponding to the variables in $\lambda_i$\footnote{Certain chunks are an exception, some chunks may have $2|\lambda_i| + 2$ literals.  Most chunks have 2 literals corresponding to 1 variable; certain chunks may have 4 literals corresponding to 1 variable.}.  We will refer to these guard locations by the corresponding literal names.  Location $x$ and $\overline{x}$ corresponding to a variable $x$ are consecutive on the chunk but either may be to the left or right of the other.  The left to right ordering of literals in a chunk $C_i$ corresponding to different variables is according to $\lambda_i$ if $i$ is even.  If $i$ is odd, the right to left ordering of the literals corresponding to different variables is according to $\lambda_i$.

Associated with each chunk $C_i$ will be a number $n_i$.  {\em In the reduction, $n_i$ guards will be needed
within chunk $C_i$ to see distinguished points in $C_i$. If less than $n_i$ guards are placed in chunk $C_i$, no matter how many guards are placed elsewhere, certain distinguished points in chunk $C_i$ will go unseen.}  For $C_0$, $\lambda_0 = \langle x_1, x_2, ..., x_n \rangle$, the literal locations are $x_1, \overline{x_1}, x_2, \overline{x_2}, ..., x_n, \overline{x_n}$ and $n_0 = n$.  To guard $C_0$ using $n_0$ guards, we will have to place exactly $n$ guards at exactly $n$ of the literal locations,
with one guard for each variable or its complement.  Note that such a placement of guards specifies an assignment to the variables.

$C_0, C_1, ..., C_k$ are used to implement the sequence $\beta$, as we now describe.  Suppose that we have added chunks $C_1, ..., C_i$ to implement steps $\beta_1, ..., \beta_j$ of $\beta$.  Let $\sigma(j)$ refer to the sequence $\sigma$ after step $\beta_j$.  By construction, chunk $C_i$ will have $\lambda_i = \sigma(j)$.  Suppose $\beta_{j+1}$ is a step in which we delete variable $x$ from $\sigma(j)$.  Chunk $C_{i+1}$ will have $\lambda_{i+1} = \sigma(j) \setminus x$.  We will have $n_i = | \lambda_i |$, and $n_{i+1} = | \lambda_{i+1} |$.  The relationship between $C_i$ and $C_{i+1}$ will be what we call a \textit{deletion}, which has the following property: to guard $C_i$ and $C_{i+1}$ using $n_i + n_{i+1}$ guards, it is necessary that we have:

\begin{enumerate}
 \item exactly $n_{i+1} = |\lambda_{i+1}|$ guards at the literals within $C_{i+1}$, one for each variable so that this corresponds to an assignment to the variables in $\lambda_{i+1}$;
 \item exactly $n_i = |\lambda_i|$ guards at the literals within $C_i$, one for each variable so that this corresponds to an assignment to the variables in $\lambda_i$;
 \item The location of the guards must be consistent for all variables except $x$: There is a guard at literal $y$ in $C_i$ if and only if there is a guard at literal $y$ in $C_{i+1}$.
\end{enumerate}

Suppose that $\beta_{j+1}$ is a clause step involving the variables $x, y$ and $z$.  This requires up to two applications of an \textit{inversion gadget} followed by a \textit{clause gadget}.  An inversion involving a variable $x$ uses three chunks $C_i$, $C_{i+1}$, and $C_{i+2}$. Its purpose
is to change the left to right ordering of literals $x$ and $\overline{x}$ in $C_{i+2}$ to be opposite of that in $C_i$.  If the relationship between $C_i, C_{i+1},$ and $C_{i+2}$ is an inversion corresponding to $x$, then $\lambda_{i+2} = \sigma(j)$ is the same as $\lambda_i$.  We have $n_i = |\lambda_i|$, 
$n_{i+1} = |\lambda_i| + 1$, and $n_{i+2} = |\lambda_{i+2}|$. To guard $C_i, C_{i+1},$ and $C_{i+2}$ using $n_i + n_{i+1} + n_{i+2}$ guards, it is necessary that we have:

\begin{enumerate}
 \item $n_{i+2} = |\lambda_{i+2}|$ guards for $C_{i+2}$, one for each variable, as above;
 \item $n_{i+1}$ guards for $C_{i+1}$;
 \item $n_i = |\lambda_i|$ guards for $C_i$, one for each variable;
 \item The location of the guards must be consistent for all variables: There is a guard at literal $y$ in $C_i$ if and only if there is a guard at literal $y$ in $C_{i+2}$.
\end{enumerate}

\begin{figure}[tpb]
\centering
\input{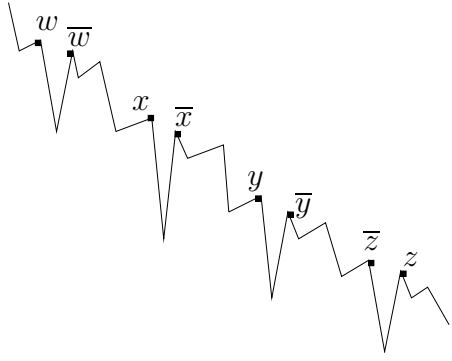}
\caption{Partial chunk showing the ordering of the variables $w, x, y,$ and $z$.}
\label{fig:clauseEx}
\end{figure}

Suppose that $\beta_{j+1}$ is a clause step involving $\overline{x} \vee y \vee \overline{z}$.  The variables $x, y$ and $z$ must occur consecutively in either left to right or right to left order in $C_i$; it must also be the case that the literal $\overline{x}$, the two literals corresponding to $y$ and the literal $\overline{z}$ occur consecutively in either left to right or right to left order in $C_i$, see Figure \ref{fig:clauseEx}.  Recall that $\beta_{j+1}$ also deletes the middle variable $y$.  By construction, chunk $C_i$ will have $\lambda_i = \sigma(j)$.  Chunk $C_{i+1}$ will have $\lambda_{i+1} = \sigma(j) \setminus y$.  We have $n_i = |\lambda_i|$ and $n_{i+1} = |\lambda_{i+1}|$. The relationship between $C_i$ and $C_{i+1}$ will be what we call a \textit{clause gadget}, which has the following property: to guard $C_i$ and $C_{i+1}$ using $n_i + n_{i+1}$ guards, it is necessary that we have:

\begin{enumerate}
 \item $n_{i+1} = |\lambda_{i+1}|$ guards for $C_{i+1}$, one for each variable;
 \item $n_i = |\lambda_i|$ guards for $C_i$, one for each variable;
 \item The location of the guards must must be consistent for all variables except $y$: There is a guard at literal $a$ in $C_i$ if and only if there is a guard at literal $a$ in $C_{i+1}$;
 \item There is a guard in $C_i$ at one of $\overline{x}$, $y$, or $\overline{z}$.
\end{enumerate}

Similar actions are done to build the chunks $C_{-1}, C_{-2}, ..., C_{-k'}$ for the $\alpha$ sequence.  Our discussion implies that chunks $C_{-k'}, ..., C_0, ..., C_k$ can be guarded with $\sum_{-k'}^k n_i$ guards if and only if we have a satisfying assignment to the planar 3SAT formula $\Phi$.  The location of the guards in chunk $C_0$ will tell us the truth value for each variable.  Our construction will be such that if $\Phi$ is satisfiable, then $\sum_{-k'}^k n_i$ guards are {\em sufficient} for seeing all distinguished points.  This will establish NP-hardness.

\section{Reduction: Gadgets}
The following subsections describe the gadgets introduced in Section 2. In Section 3.1,
we begin by describing the shape of a chunk and the location of the literals corresponding to a variable.  In
Section 3.2, we describe the basic gadget relating two chunks called the \textit{mirror gadget}.  
Subsequently, we modify the mirror gadget to obtain the deletion gadget, the inversion gadget,
and the clause gadget. We will refer to the construction of chunks $C_1, C_2, \ldots, C_k$ as ``going down''
from $C_0$, and the construction of chunks $C_{-1}, C_{-2}, \ldots, C_{-k'}$ as ``going up'' from
$C_0$.  Take an arbitrary variable $x$ in an even chunk $C_i$.  Guard locations in $C_i$ to the right of $x$ will be considered ``below'' $x$ and guard locations in $C_i$ placed to the left of $x$ will be considered ``above'' $x$.  With odd chunks, guard locations to the left are considered below and guard locations to the right are considered above.  For example, in Figure \ref{fig:clauseEx}, $w$ is above $x$; $y$ is below $x$.

\subsection{Variable Gadget}

\begin{figure}[htpb]
\centering
\input{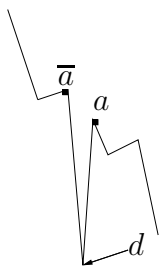}
\caption{Variable Gadget.}
\label{fig:var}
\end{figure}

The first gadget we will describe is the \textit{variable gadget}.  An example of a variable gadget for $a$ in chunk $C_i$ is shown in Figure \ref{fig:var}.  The variable gadget has a \textit{variable distinguished point}, $d$, {\em that can be seen from only two vertices: the literals $a$ and $\overline{a}$ vertices}.  The following is what we will refer to as the {\em Uniqueness Claim}:


\begin{uc}
No guard can see more than 1 variable distinguished point. 
\end{uc}
 
Because of the {\em Uniqueness Claim}, the total number of variable distinguished points provides a lower bound on the number of guards that are necessary to guard all of the distinguished points.

To see how multiple variables are placed, assume a chunk $C_i$ has $\lambda_i = \left<w,x,y,z\right>$.  Figure \ref{fig:clauseEx} shows how variable gadgets corresponding to each variable are placed within the chunk.  Chunk $C_0$ has $n$ such variable gadgets, 1 for each variable.  In Figure \ref{fig:clauseEx}, $4$ guards are required to guard the $4$ variable distinguished points because of the {\em Uniqueness Claim}.  

\paragraph{Local Summary of Variable Gadgets:}  To guard the variable distinguished point $d$ for a variable $a$ in chunk $C_i$, at least $1$ guard must be placed at the literal $a$ or $\overline{a}$ location in $C_i$.





\subsection{Mirroring}

Going down, chunks $C_i$ and $C_{i+1}$ can form what we call a mirror gadget.  Here, we will have $n_i = | \lambda_i |$, $\lambda_{i+1} = \lambda_i$, and $n_{i+1} = | \lambda_{i+1} |$.  The relationship between $C_i$ and $C_{i+1}$ will be what we call a \textit{mirroring}, which has the following property: to guard $C_i$ and $C_{i+1}$ using $n_i + n_{i+1}$ guards, it is necessary that we have:

\begin{enumerate}
 \item exactly $|\lambda_{i+1}|$ guards at the literals within $C_{i+1}$, one for each variable so that this corresponds to an assignment to the variables in $\lambda_{i+1}$;
 \item exactly $|\lambda_i|$ guards at the literals within $C_i$, one for each variable so that this corresponds to an assignment to the variables in $\lambda_i$;
 \item The location of the guards must be consistent for all variables: There is a guard at literal $y$ in $C_i$ if and only if there is a guard at literal $y$ in $C_{i+1}$.
\end{enumerate}

\begin{figure}[htpb]
\centering
\input{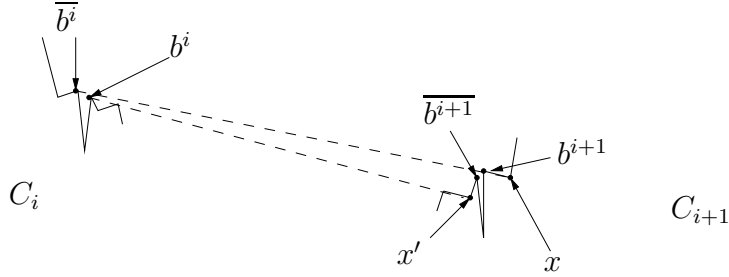}
\caption{Mirroring one variable.  Visibilities are as follows: $\overline{b^i}$ sees $x, b^{i+1}$.  $b^i$ sees $x', \overline{b^{i+1}}, b^{i+1}$. $b^{i+1}$ sees $x, \overline{b^{i+1}}, \overline{b^i}, b^i$. $\overline{b^{i+1}}$ sees $x', b^{i+1}, b^i$.}
\label{fig:mirrorUp}
\end{figure}

To describe the mirroring for $1$ variable, let us first focus on a variable gadget corresponding to a variable $b$ in chunks $C_i$ and $C_{i+1}$, see Figure \ref{fig:mirrorUp}.  We introduce the notion of \textit{mirrored distinguished points} corresponding to $b$ in $C_{i+1}$.  In Figure \ref{fig:mirrorUp}, mirrored distinguished points are $x$ and $x'$.  $b^i$ is the literal $b$ in chunk $C_i$.  $\overline{b^{i+1}}$ and $b^i$ both see our mirrored distinguished point $x'$ but neither see $x$.  $b^{i+1}$ and $\overline{b^i}$ both see $x$ but neither see $x'$.  This leads us to the following lemma:


\begin{lemma}\label{lemmaMirrorUp} For two guards to see the variable distinguished points in $C_i$ and $C_{i+1}$ corresponding to a variable $b$ and the mirrored distinguished points corresponding to variable $b$ in $C_{i+1}$, it is necessary and sufficient to place guards at the literal $b$ locations in both chunks or guards at the literal $\overline{b}$ locations in both chunks. 
\end{lemma}

\begin{proof}
Since we have two variable gadgets for $b$, the {\em Uniqueness Claim} states that two guards are necessary to guard the variable distinguished points for $b$ in $C_i$ and $C_{i+1}$.  We claim two guards are sufficient to guard the mirrored and variable distinguished points in $C_{i+1}$ and variable distinguished points in $C_i$.  We must choose one guard from $\{b^i, \overline{b^i}\}$ and one guard from $\{b^{i+1}, \overline{b^{i+1}}\}$.  If we place a guard at $b^i$, $x$ is not seen.  
Since $\overline{b^{i+1}}$ does not see $x$, we must place a guard at $b^{i+1}$.  Similar arguments can be made if we choose $\overline{b^i}$ first.
\end{proof}

Mirroring up uses a similar proof.  If a guard is placed at $b^{i+1}$, a guard must be placed at $b^i$ so that $x'$ is seen.  Similarly with $\overline{b^{i+1}}$ and $\overline{b^i}$.  

\begin{figure}[htpb]
\centering
\input{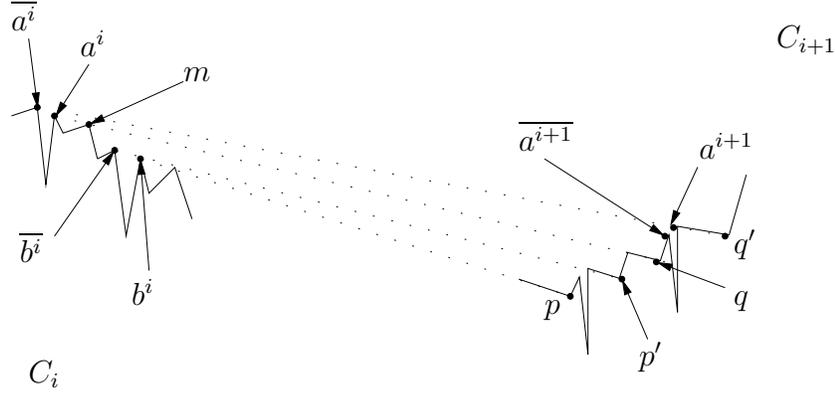}
\caption{Variable gadgets do not interfere with each other.  Important visibilities are as follows: $\overline{a^i}$ sees $q', a^{i+1}$.  $a^i$ sees $q, a^{i+1}, \overline{a^{i+1}}$.  $\overline{b^i}$ sees $p', \overline{a^{i+1}}, a^{i+1}$.  $b^i$ sees $p, a^{i+1}, \overline{a^{i+1}}$.  Note that the visibilities do not disrupt the order claim.}
\label{fig:interfere}
\end{figure}

We see in Figure \ref{fig:interfere} how variable gadgets are constructed to ensure a guard placed in one variable gadget does not see the distinguished points of a different variable gadget.  Let us say that $a^i$ and $b^i$ belong to chunk $C_i$.  $a^{i+1}$ and $b^{i+1}$ belong to chunk $C_{i+1}$.  To ensure that guards placed at a literal for one variable does not affect the mirroring of another variable, in other words $q$ should be seen by only $\overline{a^{i+1}}$ and $a^i$ and by no other guards in $C_i$ and $C_{i+1}$, similarly $q'$ should be seen by only $a^{i+1}$ and $\overline{a^i}$ and by no other guards in $C_i$ and $C_{i+1}$, the following are also true.  The line defined by $q$ and $m$ hits the terrain at point $a^i$.  Since we know $\overline{a^i}$ does not see $q$, a guard placed at $a^i$ or $\overline{a^i}$ in chunk $C_i$ will not see any of $b$'s distinguished points in $C_i$ or $C_{i+1}$.  In other words, $m$ blocks $a^i$ from seeing below variable gadget $a$ in $C_i$ and $C_{i+1}$.  In general, a guard placed at either literal for $a \in C_i$ will not see any of the mirrored or variable distinguished points of different variables below (to the right of) the variable gadget for $a \in C_i$ and also below (to the left of) the variable gadget for $a \in C_{i+1}$.

Neither $b^i$ nor $\overline{b^i}$ can see $q$ or $q'$ ensuring $b$ in $C_i$ does not affect any of $a$'s distinguished points in $C_{i+1}$.  The line defined by $q'$ and $a^{i+1}$ passes just above $a^i$ and hits the terrain just below $\overline{a^i}$.  Because of this no guard placed below this line can see $q'$.  The line defined by $q$ and $m$ hits the terrain at point $a^i$.  In Figure 5, the terrain coming out of $q$ to the left of $q$ is drawn on this line.  Therefore no guard below this line can see $q$.  In general, no guard below (to the right of) the variable gadget for $a \in C_i$ can see $q$ or $q'$ nor can any guard below (to the left of) the variable gadget for $a \in C_{i+1}$ see $q$ or $q'$.  Note that the visibilities do not disrupt the order claim.

\paragraph{Local Summary of Mirroring Gadget $C_i$--$C_{i+1}$ going down:}  To guard the variable distinguished points and mirrored distinguished points of $C_{i+1}$ and the variable distinguished points of $C_i$ with $n_i + n_{i+1}$ guards, it is necessary and sufficient to place $n_i$ guards at literals in $C_i$ and $n_{i+1}$ guards at literals in $C_{i+1}$ in a consistent way.

For mirroring up, the picture is exactly the same as above, but we proceed in the opposite direction. Note
that Lemma \ref{lemmaMirrorUp} says that if we have a guard at $a^{i+1}$, the second guard is forced to be
at $a^i$. Similarly for $\overline{a^{i+1}}$ and $\overline{a^i}$.

\paragraph{Local Summary of Mirroring Gadget $C_{i+1}$--$C_{i}$ going up:}  To guard the variable distinguished points and mirrored distinguished points of $C_{i+1}$ and the variable distinguished points of $C_i$ with $n_i + n_{i+1}$ guards, it is necessary and sufficient to place $n_i$ guards at literals in $C_i$ and $n_{i+1}$ guards at literals in $C_{i+1}$ in a consistent way.

\subsection{Deletion Gadget}

A deletion of a variable $x$ going down from chunk $C_i$ to chunk $C_{i+1}$ involves flattening out the terrain in chunk $C_{i+1}$ where the variable gadget for $x$ would have been placed.  The interested reader can read a full description of deletion in Appendix A.

\paragraph{Local Summary of Deletion Gadget $C_i$--$C_{i+1}$ going down:}  To guard the variable distinguished points and mirrored distinguished points of $C_{i+1}$ and the variable distinguished points of $C_i$ with $n_i + n_{i+1}$ guards, it is necessary and sufficient to place $n_i$ guards at literals in $C_i$ and $n_{i+1}$ guards at
literals in $C_{i+1}$ in a consistent way.

\paragraph{Local Summary of Deletion Gadget $C_{i+1}$--$C_i$--$C_{i-1}$ going up:}  To guard the variable distinguished points and mirrored distinguished points of $C_{i+1}$ and $C_i$, and the variable distinguished points of $C_{i-1}$ with $n_{i+1} + n_i + n_{i-1}$ guards, it is necessary and sufficient to place $n_{i+1}$ guards at literals in $C_{i+1}$ and $n_i$ guards at literals in $C_i$, and $n_{i-1}$ guards at literals in $C_{i-1}$ in a consistent way.

\begin{figure}[tpb]
\centering
\input{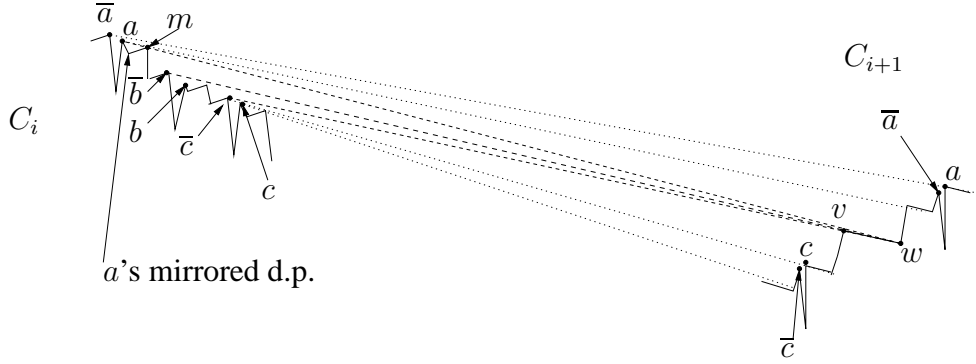}
\caption{Clause going down.}
\label{fig:clauseDown}
\end{figure}

\subsection{Downward Clause Gadget}

Let us say the clause we are constructing is $Cl_i = (a \vee \overline{b} \vee \overline{c})$, see Figure \ref{fig:clauseDown}.  We will have $\lambda_{i+1} = \lambda_i \setminus b$, $n_i = | \lambda_i |, n_{i+1} = | \lambda_{i+1} |$.  The total number of guards needed within $C_i$ and $C_{i+1}$ will be $n_i + n_{i+1}$.  We will replace the middle variable gadget $b$ in $C_{i+1}$ with our clause gadget.  In chunk $C_i$, the left to right ordering of literals if $i$ is even (right to left if $i$ is odd) must be exactly $a, \overline{b}, b, \overline{c}$.  We will assume the ordering is correct when placing a clause gadget.  Section 3.6 will show how to make a change if the ordering is incorrect.  In Figure \ref{fig:clauseDown}, $w$ is our \textit{clause distinguished point}.  We can manipulate the $b$ variable gadget in $C_i$ so that $b$ or $\overline{b}$ is blocked from seeing $w$.  In this case, $b$ is moved further down so it does not see $w$.

The original use of the $m$ point was to block a potential guard placed at the $a \in C_i$ guard location from seeing mirrored distinguished points below (to the left of) the $a$ variable gadget in $C_{i+1}$.  In this case however, we want $a \in C_i$ to see $v$.  We move our $m$ point towards the mirrored distinguished point of $a \in C_i$ so that the guard location for $a \in C_i$ sees $v$.  It should be noted that our mirroring of $a$ is not disrupted with this modification.  This modification now allows $a \in C_i$ to see $v$, which then allows $a$ to see $w$.  Note that the visibilities do not disrupt the order claim.  We also note that $\overline{a} \in C_i$ does not see $w$.  We then adjust the $vw$ line segment by moving $w$ slightly upwards so that $w$ sees $\overline{c} \in C_i$.  A ray shot from $w$ through $v$ will hit the terrain in chunk $C_i$ at point $\overline{c}$ so that $c$ in $C_i$ does not see $w$.  It should be noted that the mirroring down of $a$ and $c$ are still intact; we are still able to mirror the values of $a$ and $c$ down the terrain.  If neither $a \in C_i$ nor $\overline{b} \in C_i$ nor $\overline{c} \in C_i$ is chosen as a guard location, we require an extra guard to see $w$.  However if one of these literals is chosen to be a guard, our clause distinguished point $w$ is guarded and no extra guard is needed.



We also note that $b$ will no longer be used in any future clauses going downward.  The reduction from planar 3SAT allows us to order the clauses in a certain way to ensure that the middle variable will no longer be used in future clauses going down the terrain.  For detailed information on how the clauses are ordered, the interested reader can see Appendix C.  Because of this ordering, we can safely replace the $b$ variable gadget in $C_{i+1}$ with a clause gadget.

\paragraph{Local Summary of Clause Gadget $C_i$--$C_{i+1}$ going down:}  To guard the variable distinguished points, clause distinguished point and mirrored distinguished points of $C_{i+1}$ and the variable distinguished points of $C_i$ with $n_i + n_{i+1}$ guards, it is necessary and sufficient to place $n_i$ guards at literals in $C_i$ and $n_{i+1}$ guards at literals in $C_{i+1}$ in a consistent way.  Note that if a guard is placed at $a$ or $\overline{b}$ or $\overline{c}$ in chunk $C_i$, our clause distinguished point is seen and no additional guard is required.


\subsection{Upward Clause Gadget}

The upward clause gadget uses similar arguments as the downward clause gadget.  The interested reader can read a full description of the upward clause gadget in Appendix B. 

\paragraph{Local Summary of Clause Gadget $C_i$--$C_{i-1}$--$C_{i-2}$ going up:}  To guard the variable distinguished points, clause distinguished point and mirrored distinguished points of $C_{i-1}$, the variable distinguished points of $C_{i-2}$ and the variable and mirrored distinguished points of $C_i$ with $n_i + n_{i-1} + n_{i-2}$ guards, it is necessary and sufficient to place $n_i$ guards at literals in $C_i$ and $n_{i-1}$ guards at literals in $C_{i-1}$ and $n_{i-2}$ guards at literals in $C_{i-2}$ in a consistent way.  Note that if a guard is placed at $a$ in $C_{i-2}$ or $\overline{b}$ in $C_{i-1}$ or $\overline{c}$ in $C_{i-2}$, our clause distinguished point is seen and no additional guard is required.


\subsection{Inversion Gadget}

\begin{figure}[htpb]
\centering
\input{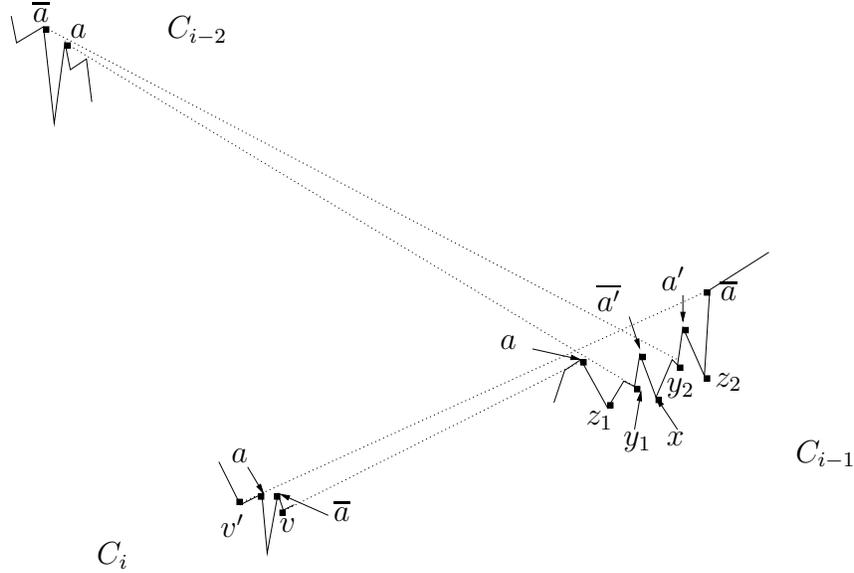}
\caption{Inverting one variable.}
\label{fig:invert}
\end{figure}

The left to right (right to left) ordering of literals becomes important when placing a clause gadget and it is possible that the literals are ``out of order.''  In a regular mirroring of variable $a$, the left to right order of $a$ and $\overline{a}$ will be the same in all even chunks, similarly with all odd chunks.  To switch the order, we make use of an inversion gadget.  Let us consider chunks $C_i, C_{i-1}, C_{i-2}$ when an inversion gadget is being placed to invert a variable, see Figure \ref{fig:invert}.  We will have $\lambda_{i-1} = \lambda_i$, $\lambda_{i-2} = \lambda_i$, $n_i = | \lambda_i |, n_{i-1} = | \lambda_{i-1} | + 1$ and $n_{i-2} = | \lambda_{i-2} |$.  The total number of guards needed will be $n_i + n_{i-1} + n_{i-2}$.  The $a$ literal in $C_{i-2}$ is to the right of $\overline{a}$ in $C_{i-2}$.  Using the inversion gadget in $C_{i-1}$, we can swap the left to right ordering of the $a$ and $\overline{a}$ literal so that $a$ in $C_i$ is to the left of $\overline{a}$ in $C_i$.


\subsubsection{Inverting Down}

In Figure \ref{fig:invert}, the variable gadget for $a$ in chunk $C_{i-1}$ is replaced with an inversion gadget.  The inversion gadget adds two literal locations for variable $a$ in $C_{i-1}$, namely $a'$ and $\overline{a'}$.  The variable and mirrored distinguished points of $a \in C_{i-1}$ are being replaced with five inversion distinguished points.  These five \textit{inversion distinguished points} are: $(x, z_1, y_1, y_2, z_2)$.  $y_1$ and $y_2$ should be thought of as ``mirrored distinguished points'' since they are seen by guards inside chunk $C_{i-1}$ and by the $a$ and $\overline{a}$ literal guard locations in chunk $C_{i-2}$.  $z_1$ and $z_2$ are the replacement ``variable distinguished points.''  They are replacement variable distinguished points in the sense that no guard outside of the inversion gadget for $a$ in $C_{i-1}$ can see them.  More importantly, $z_1$ and $z_2$ are considered replacement ``variable distinguished points'' because they obey the {\em Uniqueness Claim}.  $z_1$ is only seen by $a \in C_{i-1}$ and $\overline{a'} \in C_{i-1}$.  $z_2$ is only seen by $\overline{a} \in C_{i-1}$ and $a' \in C_{i-1}$.  The $x$ point is the special inversion distinguished point that allows the inversion to take place.  

The important visibilities are given here: $a \in C_{i-1}$ sees $z_1$ and $v$; $\overline{a'} \in C_{i-1}$ sees $z_1$, $y_1$ and $x$;  $a' \in C_{i-1}$ sees $x$, $y_2$ and $z_2$; $\overline{a} \in C_{i-1}$ sees $z_2$ and $v'$; $a \in C_{i-2}$ sees $y_1$; and $\overline{a} \in C_{i-2}$ sees $y_2$.  Although not entirely obvious from the Figure, it's important to note a ray shot from $\overline{a} \in C_{i-1}$ through $a \in C_{i-1}$ hits the terrain to the left of $v \in C_i$.  These visibilities do not disrupt the order claim.

Because of the {\em Uniqueness Claim}, it is necessary that we place $4$ guards to see the variable distinguished points of $C_i, C_{i-1}$ and $C_{i-2}$.  If we place the $4$ guards in a consistent manner, the mirrored distinguished points of $C_i$ and the inversion distinguished points of $C_{i-1}$ will also be seen.  We will only concern ourselves with the inversion of $a$ and ignore the other variables being mirrored.  The other variables are being mirrored without consequence.  We will assume we already have a guard at $a \in C_{i-2}$ or $\overline{a} \in C_{i-2}$.  Because of the {\em Uniqueness Claim}, it is necessary that we place $3$ guards to see the remaining ``variable distinguished points'' of $z_1$ and $z_2$ and also the variable distinguished point of $a \in C_i$.  If we place the remaining $3$ guards in a consistent manner, the remaining distinguished points of $y_1, y_2, x, v$ and $v'$ will also be seen.  

We note that we need to place one guard at $a \in C_i$ or $\overline{a} \in C_i$ to see the variable distinguished point for $a \in C_i$.  This leaves two guards to be placed to see the remaining ``variable distinguished points'' of $z_1$ and $z_2$.

Using the example in Figure \ref{fig:invert}, let us say $a \in C_{i-2}$ was chosen to be a guard.  We know at least $1$ guard must be placed at $a \in C_i$ or $\overline{a} \in C_i$ to see the variable distinguished point of $a$ in $C_i$ leaving $2$ guards to see the unguarded inversion distinguished points: $y_2, x, z_1$ and $z_2$.  Let us first consider who can guard $y_2$.  The only 2 guards that see $y_2$ are $a' \in C_{i-1}$ and $\overline{a} \in C_{i-2}$.  If we place a guard at $\overline{a} \in C_{i-2}$, one of the ``variable distinguished points'' of $z_1$ or $z_2$ will go unseen.  Therefore we must choose to place our guard at $a' \in C_{i-1}$.

We have one guard left to place in the inversion gadget that must see $z_1$.  In order to see both $v$ and $v'$, we must place our guard at $a \in C_{i-1}$.  The only other choice is $\overline{a'} \in C_{i-1}$ but this guard does not see $v$ or $v'$.  Placing a guard at $a \in C_{i-1}$ leaves $v'$ and the variable distinguished point of $a \in C_i$ unguarded.  $a \in C_i$ is chosen to be a guard and the inversion is complete.  Similar arguments are made showing that if $\overline{a} \in C_{i-2}$ is chosen, then $\overline{a'} \in C_{i-1}$, $\overline{a} \in C_{i-1}$ and $\overline{a} \in C_{i-1}$ must be chosen.

\paragraph{Local Summary of Inversion Gadget $C_{i-2}$--$C_{i-1}$--$C_i$ going down:}  To guard the variable distinguished points and mirrored distinguished points of $C_i$, the variable, mirrored and inversion distinguished points of $C_{i-1}$ and the variable distinguished points of $C_{i-2}$ with $n_i + n_{i-1} + n_{i-2}$ guards, it is necessary and sufficient to place $n_i$ guards at literals in $C_i$, $n_{i-1}$ guards in $C_{i-1}$, and $n_{i-2}$ guards at literals in $C_{i-2}$ in a consistent way.  If variable $a$ is being inverted, the left to right ordering of the literals $a$ and $\overline{a}$ in $C_{i-2}$ are opposite of that in $C_i$.

\subsubsection{Inverting Up}

Similar arguments are used to show the inversion going up.  We will have $\lambda_{i-1} = \lambda_{i-2}$, $\lambda_i = \lambda_{i-2}$, $n_i = | \lambda_i |, n_{i-1} = | \lambda_{i-1} | + 1$ and $n_{i-2} = | \lambda_{i-2} |$.  The total number of guards needed will be $n_i + n_{i-1} + n_{i-2}$.  


\paragraph{Local Summary of Inversion Gadget $C_i$--$C_{i-1}$--$C_{i-2}$ going down:}  To guard the variable distinguished points and mirrored distinguished points of $C_i$, the variable, mirrored and inversion distinguished points of $C_{i-1}$ and the variable distinguished points of $C_{i-2}$ with $n_i + n_{i-1} + n_{i-2}$ guards, it is necessary and sufficient to place $n_i$ guards at literals in $C_i$, $n_{i-1}$ guards in $C_{i-1}$, and $n_{i-2}$ guards at literals in $C_{i-2}$ in a consistent way.  If variable $a$ is being inverted, the left to right ordering of the literals $a$ and $\overline{a}$ in $C_{i-2}$ are opposite of that in $C_i$.




\subsection{Local vs Global View of Gadgets}

Having completed the construction, we see that for every chunk $C_i$ we need $n_i$ points placed within the chunk just to guard the variable distinguished points within the chunk.  This is given by the {\em Uniqueness Claim}.  We now observe that the local summary of any gadget holds good in a global sense, that is, it is independent of how guards are placed in chunks outside this gadget.  We illustrate this by summarizing a mirror gadget using chunks $C_i$ and $C_{i+1}$ going down.  The reader may find it useful to compare with the local summary in Section 3.2.

\paragraph{Global Summary of Mirroring Gadget $C_i$--$C_{i+1}$ going down:}  To guard the variable distinguished points and mirrored distinguished points of $C_{i+1}$ and the variable distinguished points of $C_i$ with $n_i$ guards in $C_i$ and $n_{i+1}$ guards in $C_{i+1}$, it is necessary to place $n_i$ guards at literals in $C_i$ and $n_{i+1}$ guards at literals in $C_{i+1}$ in a consistent way.  This necessity holds good for {\em any} placement of guards in locations outside $C_i$ and $C_{i+1}$.  The local sufficiency condition obviously holds good for any placement of guards outside $C_i$ and $C_{i+1}$.

What this stronger condition means, in the context of Figure \ref{fig:interfere}, is that $\overline{a^{i+1}}$ and $a^i$ are the only guard locations that see $q$ among {\em all possible} guard locations on the terrain.  Similarly for points $q', p$ and $p'$.  The argument for why this holds is the same as the one made for guard locations within $C_i$ and $C_{i+1}$.  Having completed the entire construction, we are only now in a position to state this global property.  The ``necessary'' parts of each of the gadgets are similarly modified to hold in a global sense.

\subsection{Putting it all Together}
Each chunk $C_i$ in our construction needs $n_i$ guards within it.  Because of the {\em Uniqueness Claim}, the terrain we construct needs at least $\sum_{-k'}^k n_i$ guards just to see all of the variable distinguished points. Our construction ensures that if the distinguished points can be seen by $\sum_{-k'}^k n_i$ guards, then the input formula must be satisfiable. In particular, the assignment for the variables chosen by the $n_0$ points in chunk $C_0$ must be consistently mirrored to all chunks and the clause distinguished points must be seen.  If the input formula is satisfiable, picking a satisfying assignment and propagating it through our gadgets in the natural way results in a set of $\sum_{-k'}^k n_i$ guards that see all of the distinguished points. Thus the proof of NP-hardness is thus completed.

\begin{theorem}
Discrete terrain guarding is NP-hard.
\end{theorem}

\subsection{Continuous Version}
Using the same construction, it can be shown that the continuous version of the terrain guarding problem is also NP-hard.  We argue that the entire terrain can be seen by $\sum_{-k'}^k n_i$ guards if and only if the input formula is satisfiable.  The {\em Uniqueness Claim} holds true despite guards being able to be placed anywhere on the terrain; since there are $n_i$ variable distinguished points in chunk $i$, it follows that $\sum_{-k'}^k n_i$ are necessary for seeing the entire terrain.  We now argue that if $\sum_{-k'}^k n_i$ see the entire terrain, they can be assumed to be in guard locations from the earlier reduction.  From this, it follows that the formula is satisfiable.

Referring to Figure \ref{fig:var}, the only potential guards that see $d$ are points on a line segment $\overline{a}d$ and points on a line segment $da$.  Let's say we pick a guard $g$ on the line segment $\overline{a}d$.  $\overline{a}$ will see every point that $g$ does.  If we choose $g$ as our guard, we can simply move our guard to $\overline{a}$ without any loss of visibility.  Similar arguments can be said about $a$ and the line segment $da$.  Therefore, we assume that any guard placed in the sub-terrain $\overline{a}da$ is either at $a$ or $\overline{a}$. In particular, the only potential guards for $d$ are $a$ and $\overline{a}$.  Simiarly arguments are made for the ``variable distinguished points'' in the inversion gadget.  Therefore the {\em Uniqueness Claim} holds true in the continuous version; in other words the lower bound on the number of guards necessary to guard the variable distinguished points is the same in the continuous version as in the discrete version.  

If the formula is satisfiable, $\sum_{-k'}^k n_i$ guards will see the entire terrain if the guards are placed in satisfying locations.  Clearly the distinguished points are all seen.  It can be shown that the terrain within the chunks is seen.  The ``empty space'' outside of the chunks is also seen.  For any 2 chunks $C_i$ and $C_{i-2}$ where $i= k, k-1, k-2, ..., 0, -1, ..., -k'+3, -k'+2$, any guard in chunk $C_{i-1}$ will see the ``empty space'' between $C_i$ and $C_{i-2}$ because of the order claim.  The ``empty space'' above chunk $C_{-k'+1}$ is seen by the guard placed at the literal for the last deleted variable in chunk $C_{k'}$.  The ``top'' of the terrain is drawn in such a way that a guard placed for the last variable being deleted while ``going up'' will see the highest part of the terrain in chunk $C_{k'}$.  As for the ``bottom'' of the terrain, the terrain can be slightly modified between chunk $C_{k-1}$ to $C_k$ so that the terrain connecting those two chunks is seen by the only two remaining literals in chunk $C_{k-1}$.  The entire terrain is thus seen.

\section{Conclusion and Future Work}
We have shown that terrain guarding is NP-hard. With the PTAS for terrain guarding given by Gibson et al. \cite{G09}, this essentially resolves the approximability of the problem. The biggest remaining question regarding the complexity of terrain guarding is whether or not it is fixed-parameter tractable.

\section*{Acknowledgments}
We would like to thank Kasturi Varadarajan and Bengt Nilsson for their valuable comments, discussions and suggestions.

\newpage 

\appendix
\section*{Appendix A: Deletion Gadget}
\begin{figure}[htpb]
\centering
\input{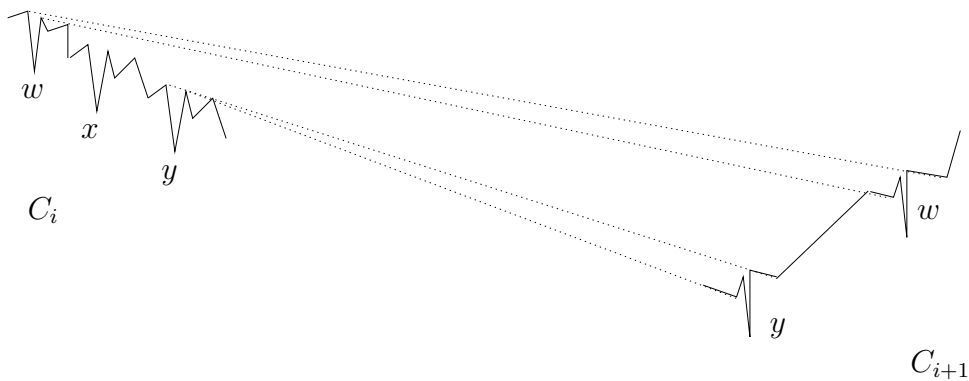}
\caption{Deleting a variable when mirroring down.}
\label{fig:delete}
\end{figure}

Let us consider chunks $C_i$ and $C_{i+1}$ going down when a deletion gadget is being placed to delete variable $x$.  The total number of guards needed will be $n_i + n_{i+1}$ where $n_i = | \lambda_i |$ and $n_{i+1} = | \lambda_{i+1} |$.  The list of variables in $C_{i+1}$, $\lambda_{i+1} = \lambda_i \setminus x$.  We replace the variable gadget for $x$ in $C_{i+1}$ with a flat surface as seen in Figure \ref{fig:delete}.

\begin{figure}[htpb]
\centering
\input{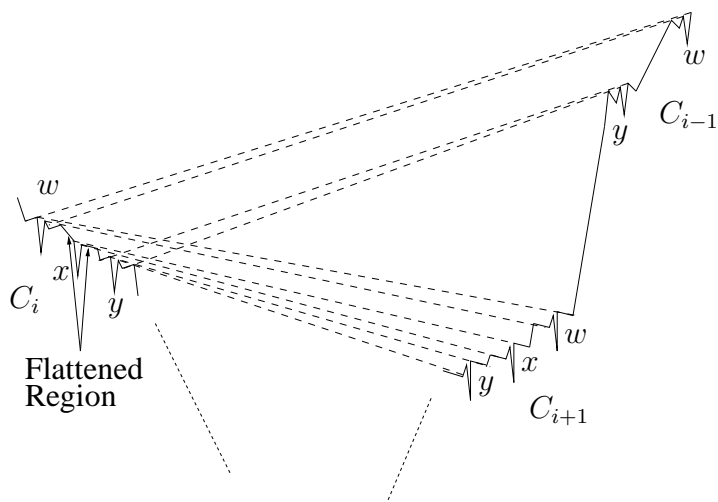}
\caption{Deleting a variable when mirroring up.}
\label{fig:deleteUp}
\end{figure}

Going up, we need three chunks $C_{i+1}$, $C_i,$ and $C_{i-1}$ to construct a deletion gadget for
deleting variable $x$. We will have $\lambda_i = \lambda_{i+1}$, $\lambda_{i-1} =
\lambda_{i+1} \setminus x$, $n_{i+1} = | \lambda_{i+1} |, n_i = | \lambda_i |$ and $n_{i-1} = | \lambda_{i-1} |$.  The total number of guards needed will be $n_{i+1} + n_i + n_{i-1}$.  We flatten out the mirrored distinguished points of variable gadget $x$ in $C_i$ as seen in Figure \ref{fig:deleteUp}.  The mirrored distinguished points were there to help us mirror $x$ up the terrain.  However, $x$ is no longer needed so the mirrored distinguished points can go away as shown.

\newpage

\section*{Appendix B: Upward Clause Gadget}
\begin{figure}[tpb]
\centering
\input{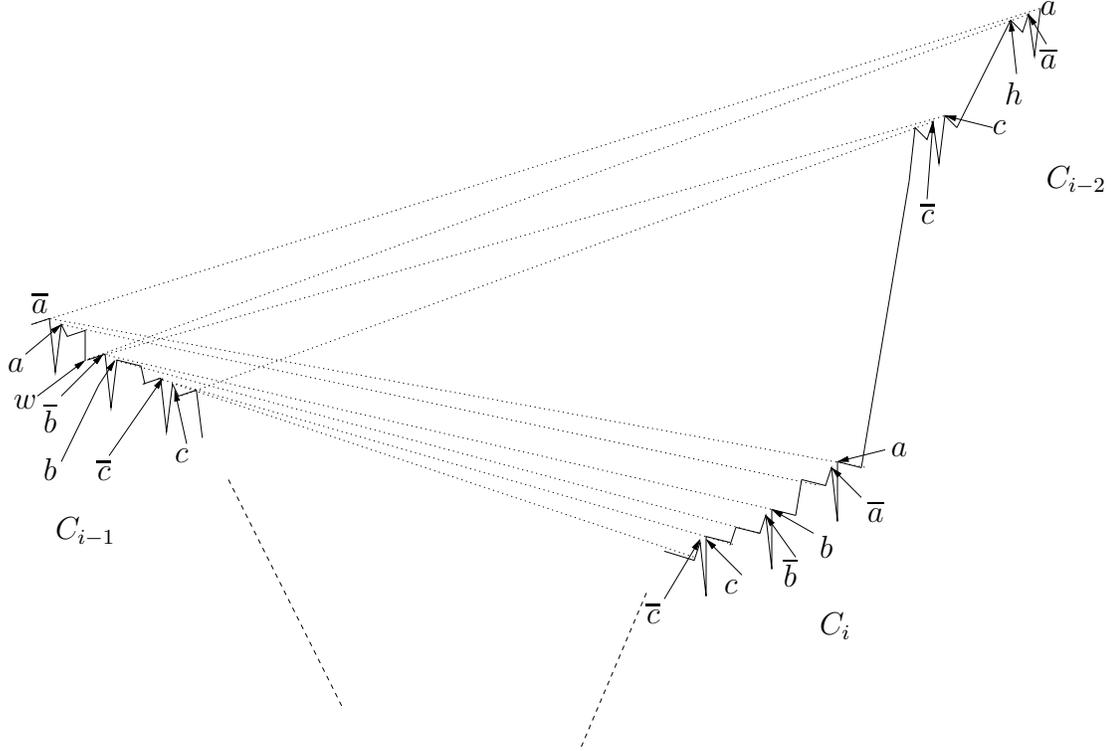}
\caption{Clause going up.}
\label{fig:clauseUp}
\end{figure}

The clause gadget going up is done similarly to mirroring variables upward with a few small changes.  We will have $\lambda_{i-1} = \lambda_i, \lambda_{i-2} = \lambda_i \setminus b$, $n_i = | \lambda_i |, n_{i-1} = | \lambda_{i-1} |$ and $n_{i-2} = |\lambda_{i-2}|$.  The total number of guards needed will be $n_i + n_{i-1} + n_{i-2}$.  We replace the highest (in this case leftmost) mirroring distinguished point of $b \in C_{i-1}$ with a clause distinguished point $w$.  We flatten out the other mirroring distinguished point for $b \in C_{i-1}$ similar to the deletion gadget.

The original purpose of our $h$ point was to ensure that $\overline{a} \in C_{i-2}$ did not affect variables below (to the right of) $a \in C_{i-1}$.  However, we want $\overline{a} \in C_{i-2}$ to see $w$ so $h$ is adjusted accordingly for this.  It should be noted that $a \in C_{i-2}$ does not see $w$.  We also allow $c \in C_{i-2}$ to see $w$.  The original reason to not allow this was so $c$ would not affect the mirroring of $b$.  This is no longer the case so we allow $c \in C_{i-2}$ to see $w$.  $w$ is adjusted accordingly so that a ray shot from $w$ through $\overline{b} \in C_{i-1}$ sees $c \in C_{i-2}$.  We now have only three variables that can see $w$: $\overline{b}  \in C_{i-1}$, $\overline{a} \in C_{i-2}$, and $c \in C_{i-2}$.  We note that $b$ can safely disappear as it will not be needed in any other clauses going upwards because of the ordering of the clauses.  See Appendix $C$ for a detailed explanation of why $b$ can be removed.

\newpage

\section*{Appendix C: Reduction Example}
\begin{figure}[tpb]
\centering
\input{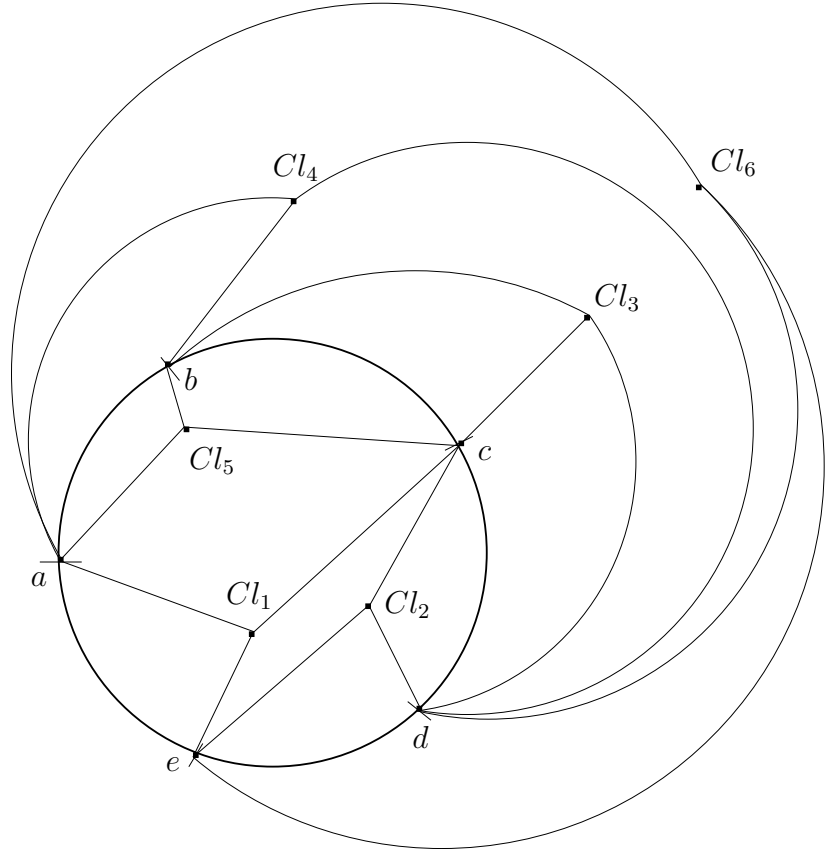}
\caption{Planar 3SAT Example.}
\label{fig:planarNew}
\end{figure}

The reduction from planar 3SAT is done in the following way.  The reduction given in this section combines several steps when visibility is not affected in an effort to minimize the number of figures needed.  For example, several variables might be deleted in one chunk where the specification calls for only one deletion per chunk.  

In Figure \ref{fig:planarNew} we have an instance of planar 3SAT.  There are clauses outside the variable cycle and clauses inside the variable cycle.  A clause $Cl_i$ is connected to 3 distinct variables.  For example, $Cl_2 = (\overline{c} \vee \overline{d} \vee e)$.  We arbitrarily pick a variable to be the lowest indexed variable and work clockwise around the variable cycle in increasing order.  In the example in Figure \ref{fig:planarNew} we choose $a$ to be the lowest index variable.  Our ordering of variables is $a<b<c<d<e$.  This indexing of variables also gives us the ordering of variable gadgets in chunks.  In even chunks, $a$ will be the leftmost variable gadget, $b$ will be the next leftmost, followed by $c$ and so on.  In odd chunks, $a$ will be the rightmost variable gadget, $b$ will be the next rightmost, followed by $c$ and so on.  Chunk $C_0$ is shown in Figure \ref{fig:ex1}.  The $interval$ of a clause $Cl_i$ is denoted $I(Cl_i)$.  The interval of a clause is defined as the span from the lowest index variable in $Cl_i$ to the highest indexed variable in $Cl_i$.  From the example, $I(Cl_3) = (b,d)$.  

We will focus on clauses outside of the variable cycle.  We assume that each clause has $3$ distinct variables.  Because of this and because our graph is planar, every clause outside the variable cycle has a unique interval.  Take two clauses outside the cycle $Cl_i$ and $Cl_j$.  We know the intervals are distinct.  Because of planarity, either the intervals $I(Cl_i)$ and $I(Cl_j)$ have disjoint interiors or one of the intervals is properly contained in the other.  If $I(Cl_i)$ is properly contained within $I(Cl_j)$, we say that clause $Cl_i < Cl_j$.  We therefore have a partial ordering of the clauses.  With this partial ordering we construct a valid total ordering of all of the clauses both inside the variable cycle and outside the variable cycle.  The ordering of clauses outside the cycle are used when placing clauses going up the terrain.  Similarly, intervals $I(Cl_i)$ and $I(Cl_j)$ for clauses inside the variable ring either have disjoint interiors or $I(Cl_i)$ is properly contained within $I(Cl_j)$.

Let us consider the ordering of clauses outside of the cycle, call this ordering $\Gamma$.  It is because of this ordering that we can delete the ``middle'' variable from the terrain when we place our clause gadget.  The middle variable is defined as the variable that is not an endpoint of the interval.  For example, if $Cl_i = (c \vee h \vee r)$ and $I(Cl_i) = (c,r)$, our middle variable is $h$.  Let us take the first clause in $\Gamma$, call this clause $Cl_i$.  We know that for every other clause $Cl_j \in \Gamma, Cl_j \nless Cl_i$.  Since we are placing the smallest $I(Cl_i)$ first, based on our partial ordering, we know there are no clauses less than $Cl_i$.  Since intervals do not overlap because of planarity, no other clause $Cl_j \in \Gamma$ will use the middle variable of $Cl_i$.  Before we can place a clause gadget for $Cl_i$, there may be unused variables in the span of $I(Cl_i)$.  We must first delete these unused variables before generating a clause gadget for $Cl_i$.

As an example, consider Figure \ref{fig:planarNew}.  We have three clauses on the outside of the cycle and three clauses on the inside of the cycle.  $Cl_1 = (a \vee \overline{c} \vee \overline{e}), Cl_2 = (\overline{c} \vee \overline{d} \vee e), Cl_3 = (b \vee c \vee d), Cl_4 = (a \vee b \vee d), Cl_5 = (a \vee b \vee c), Cl_6 = (\overline{a} \vee d \vee e)$.  The intervals of each of the clauses are $I(Cl_1) = (a,e), I(Cl_2) = (c,e), I(Cl_3) = (b,d), I(Cl_4) = (a,d), I(Cl_5) = (a,c), I(Cl_6) = (a,e)$.  A partial ordering of the clauses outside the cycle is $Cl_3 < Cl_4$ and $Cl_4 < Cl_6$ and $Cl_3 < Cl_6$.  A possible total ordering for clauses outside the ring is then $\left<Cl_3, Cl_4, Cl_6\right>$.  A partial ordering of clauses inside the cycle is $Cl_2 < Cl_1$ and $Cl_5 < Cl_1$.  A possible total ordering for clauses inside the ring is $\left<Cl_2, Cl_5, Cl_1\right>$ or $\left<Cl_5, Cl_2, Cl_1\right>$.

\begin{figure}[tpb]
\centering
\input{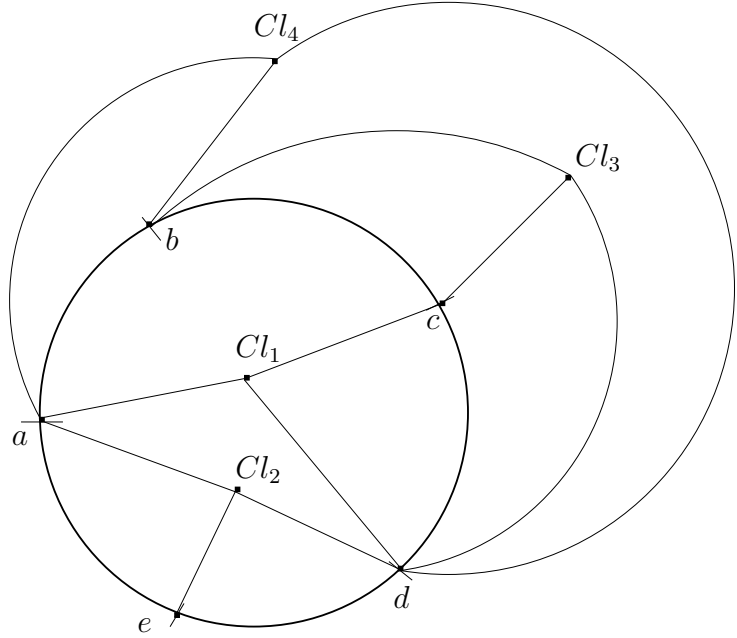}
\caption{Planar 3SAT Example.}
\label{fig:planarOld}
\end{figure}

The remainder of the example will use the planar 3SAT example shown in Figure \ref{fig:planarOld}.  The clauses for the example are defined as $Cl_1 = (a \vee \overline{c} \vee \overline{d}), Cl_2 = (\overline{a} \vee \overline{d} \vee e), Cl_3 = (b \vee c \vee d), Cl_4 = (a \vee b \vee d)$.

\begin{figure}[tpb]
\centering
\input{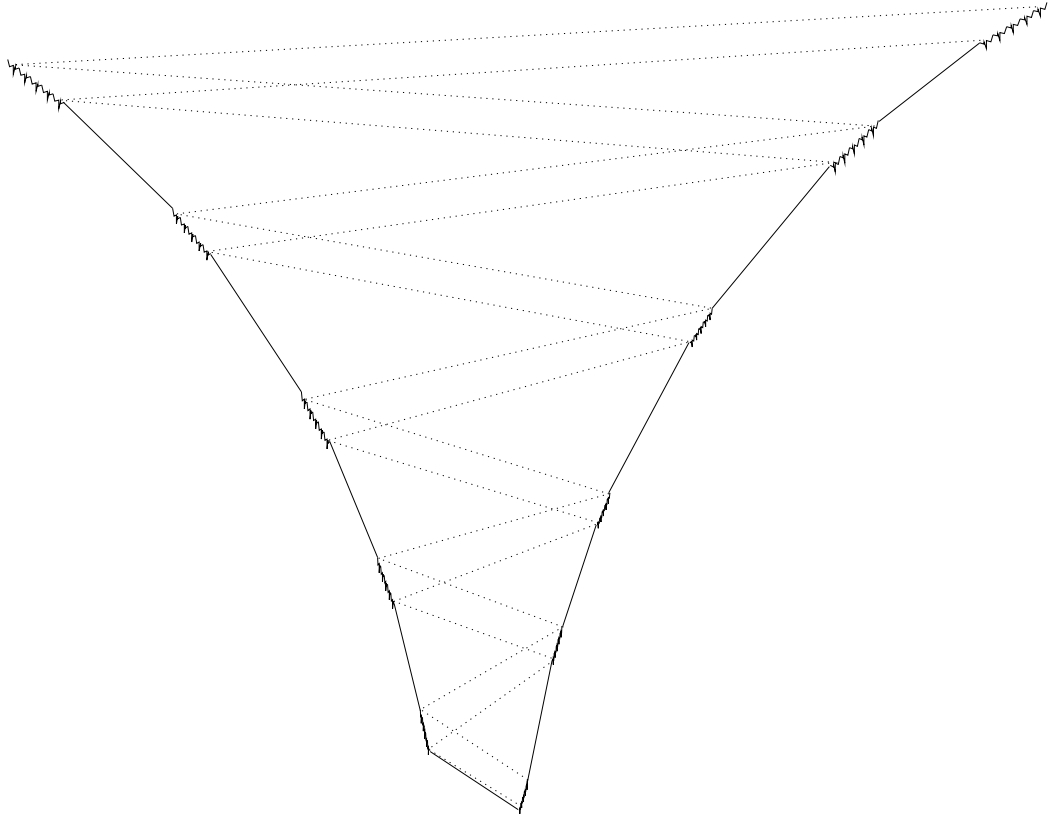}
\caption{The entire terrain.}
\label{fig:ex}
\end{figure}

\begin{figure}[tpb]
\centering
\input{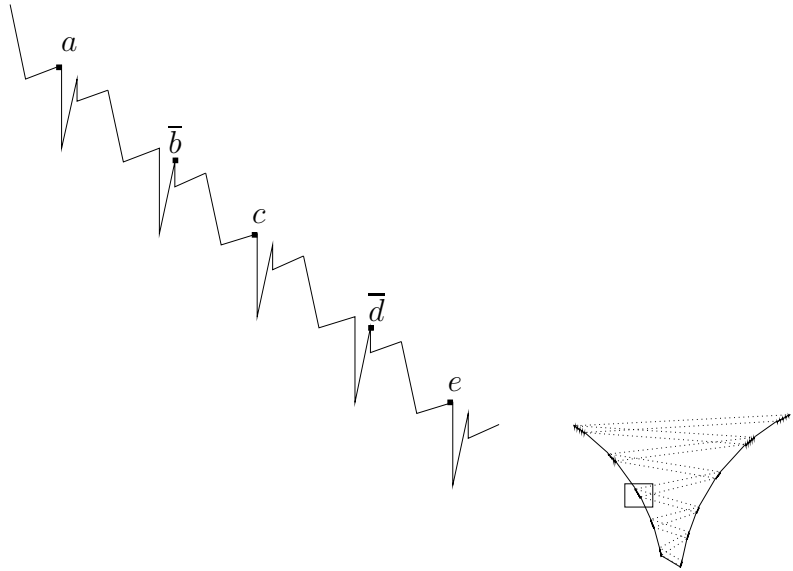}
\caption{Chunk $C_0$ which contains variable gadgets for all variables on the variable cycle.}
\label{fig:ex1}
\end{figure}

An overview of the entire terrain is shown in Figure \ref{fig:ex}.  In each subsequent figure, we will show the specific part of the terrain we are describing along with a smaller version of the entire terrain to give reference to where we are on the terrain.

\begin{figure}[tpb]
\centering
\input{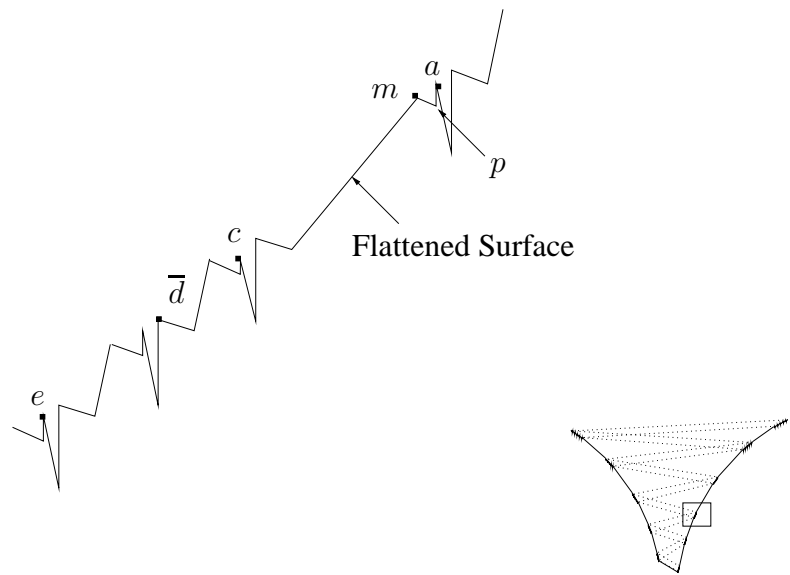}
\caption{Chunk $C_1$ which places a deletion gadget for variable $b$.}
\label{fig:ex2}
\end{figure}

Figure \ref{fig:ex1} shows the details of chunk $C_0$.  Starting at chunk $C_0$ we will work our way downward.  The figures place small rectangles for literals chosen as guard locations.  In our example, we have chosen to place guards at the literal locations $a, \overline{b}, c, \overline{d},$ and $e$.  We choose clause $Cl_1$ as the first clause placed on the terrain going downward from our total ordering obtained before.

\begin{figure}[tpb]
\centering
\input{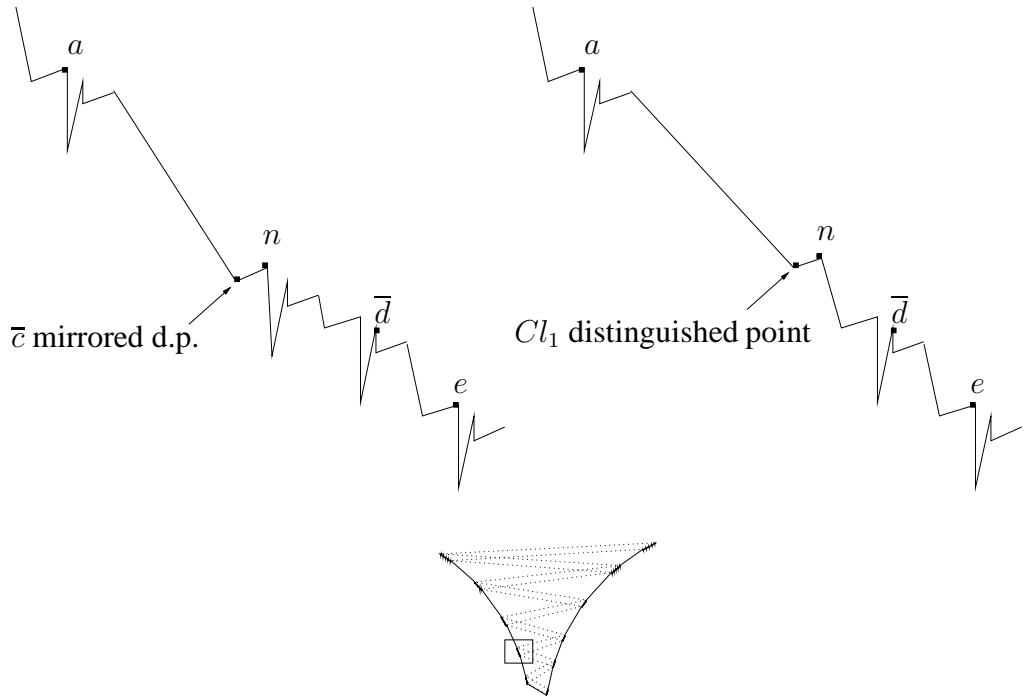}
\caption{Chunk $C_2$ which places a clause gadget for clause $Cl_1$.}
\label{fig:ex3}
\end{figure}

Before we can place a gadget for clause $Cl_1$ on the terrain, we must delete the $b$ variable since $I(Cl_1) = (a,d)$ and we only use the $a, c$ and $d$ variables.  This deletion is done in chunk $C_1$ as shown in Figure \ref{fig:ex2}.  In this figure, the location where the $b$ variable gadget would be in $C_1$ is replaced by a flat surface.

\begin{figure}[tpb]
\centering
\input{bigExample4.pstex_t}
\caption{Chunk $C_3$ which places an inversion gadget for variable $d$.}
\label{fig:ex4}
\end{figure}

We are now ready to place a clause gadget and this is shown in Figure \ref{fig:ex3}.  In this figure, the $c$ variable gadget, as seen in the left in Figure \ref{fig:ex3}, is replaced with a clause gadget for clause $Cl_1$, as seen in the right in Figure \ref{fig:ex3}.  If we were only doing a mirroring, a ray shot from the $\overline{c}$ mirrored distinguished point in chunk $C_2$ through $n$ would hit the $\overline{c}$ literal location in chunk $C_1$.  However, since we are replacing the $c$ variable gadget with a clause gadget, we want two other literals to see into $c$'s variable region, which now contains the $Cl_1$ distinguished point.  Those literal locations being $\overline{d}$ in $C_1$ and $a$ in $C_1$.  The $Cl_1$ distinguished point is adjusted upward accordingly so that a ray shot from $Cl_1$'s distinguished point through $n$ would hit the guard location for $\overline{d} \in C_1$.  Referring back to Figure \ref{fig:ex2}, we move the $m$ towards $p$ so that $a \in C_1$ sees $Cl_1$'s distinguished point in chunk $C_2$.  Recall that $m$ was originally there to keep $a$ from seeing into variable gadgets to the right of $a$ in chunk $C_2$.  We also move $c \in C_1$ down so that it does not see $Cl_1$'s distinguished point.  There are now 3 literal locations that can see $Cl_1$'s distinguished point, namely $a \in C_1, \overline{c} \in C_1,$ and $\overline{d} \in C_1$.  If any of these literals have a guard at their location, $Cl_1$'s distinguished point is seen, in other words, $Cl_1$ is satisfied.  It is also important to note that none of these modifications affect the mirroring of variables $a$ and $d$ or any other variable.

\begin{figure}[tpb]
\centering
\input{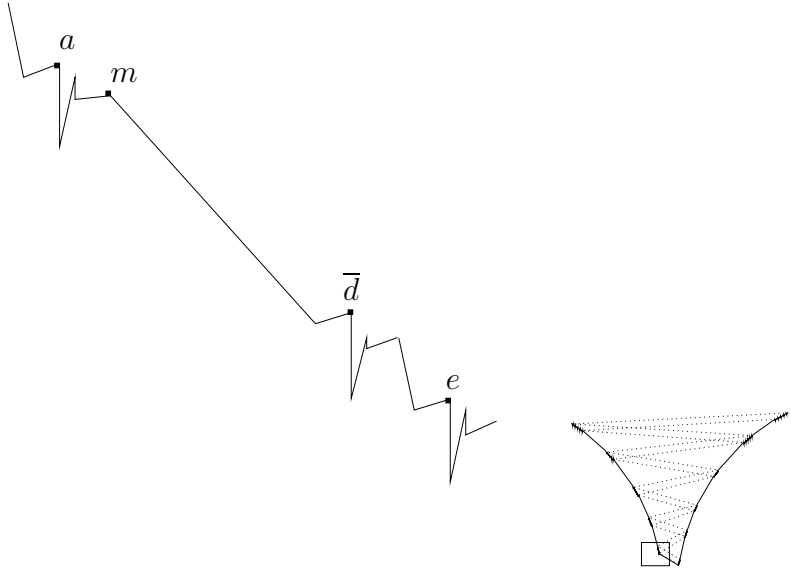}
\caption{Chunk $C_4$.}
\label{fig:ex5}
\end{figure}

The next clause that is placed is $Cl_2 = (\overline{a} \vee \overline{d} \vee e)$. Literals $\overline{a}$ and $e$ are in the correct location in chunk $C_2$ but $\overline{d}$ is not.  Therefore, the $d$ variable must be inverted before we can place the next clause gadget.  We invert $d$ in chunk $C_3$ as shown in Figure \ref{fig:ex4}.  After the inversion gadget is placed in chunk $C_3$, the ordering of the $d$ and $\overline{d}$ literals in chunk $C_4$ is correct.  This is shown in Figure \ref{fig:ex5}.  

\begin{figure}[tpb]
\centering
\input{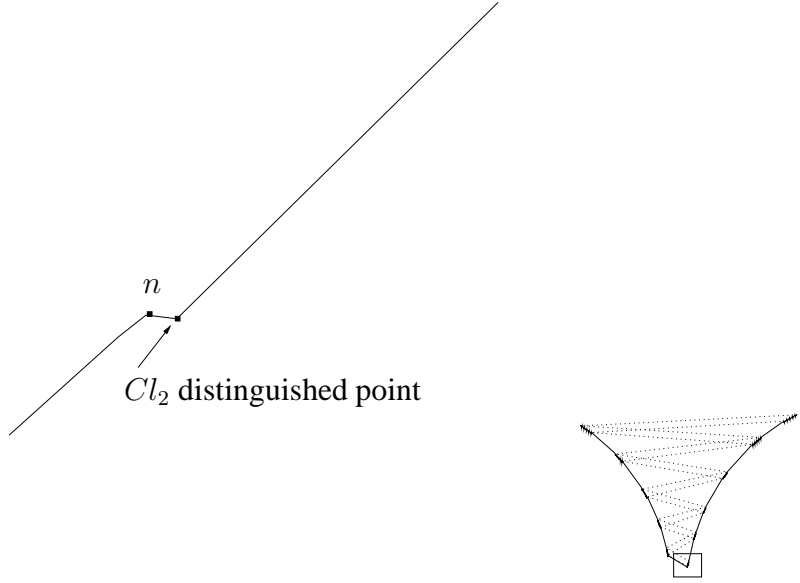}
\caption{Chunk $C_5$ which places a clause gadget for clause $Cl_2$ and a deletion gadget for variable $a$ and $e$.}
\label{fig:ex6}
\end{figure}

Once the literals are in the correct order, we can place our clause gadget for $Cl_2$.  This is done in chunk $C_5$ as shown in Figure \ref{fig:ex6}.  In this figure, the $d$ variable gadget is replaced with a clause gadget for clause $Cl_2$.  The purpose of the $n$ point in Figure \ref{fig:ex6} was to ensure variable gadgets below $d$ in chunk $C_4$ did not see into $d$'s variable gadget.  In this case, we want 1 other literal to see into $d$'s variable region, which is now our $Cl_2$ distinguished point, that point being $e \in C_4$.  The $Cl_2$ distinguished point is adjusted up accordingly.  The literal $d \in C_4$ is also adjusted down accordingly.  Referring back to Figure \ref{fig:ex5}, we adjust the $m$ point so that $\overline{a} \in C_4$ sees $Cl_2$'s distinguished point.  Recall that $m$ was originally there to keep $a$ from seeing into different variable gadgets in chunk $C_5$.  There are now 3 literal locations that can see $Cl_2$'s distinguished point, namely $\overline{a} \in C_4, \overline{d} \in C_4,$ and $e \in C_4$.  If any of these literals have a guard at their location, $Cl_2$'s distinguished point is seen and the clause is satisfied.

As said in the beginning of this section, we are combining several deletions in chunk $C_5$ to save space.  We show the deletion of the $a$ and $e$ variable in Figure \ref{fig:ex6}.  The variable gadgets are simply replaced with flat surfaces.

This ends our reduction going downward and we count how many guards are necessary going downward.  We see that each variable gadget requires a guard to be placed at one of the literal points for that particular variable gadget.  No other point on the terrain sees these variable distinguished points so a guard is required to be placed at a literal location for each variable gadget.  We also add 1 extra guard for each inversion gadget that was placed.  Recall that an inversion requires 1 extra guard because the inversion gadget has 2 ``variable distinguished points''.  We count the number of variable gadgets in each chunk $C_0, C_1, C_2, C_3, C_4,$ and $C_5$ and end up with $5+4+3+(3+1)+3+0 = 19$.

\begin{figure}[tpb]
\centering
\input{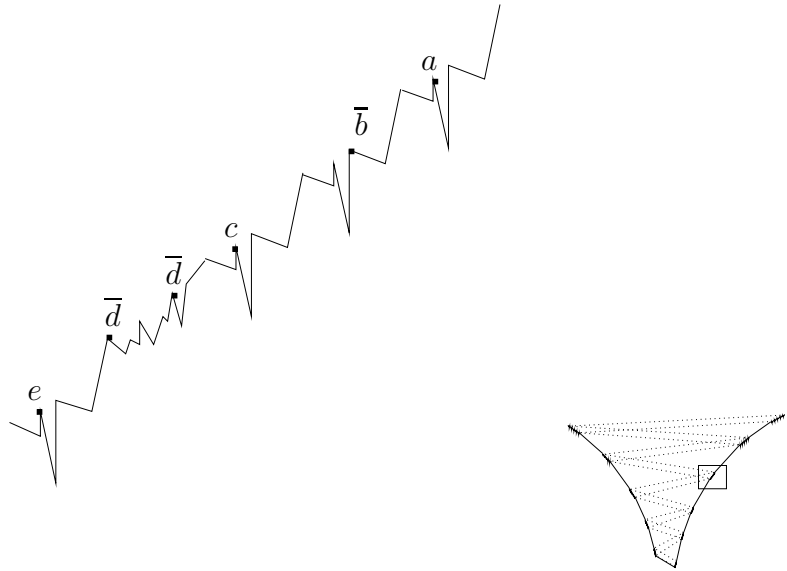}
\caption{Chunk $C_{-1}$ which places an inversion gadget for variable $d$.}
\label{fig:ex7}
\end{figure}

We now consider placing gadgets going ``up'' the terrain.  We wish to place a gadget for clause $Cl_3$ on our terrain but we must invert variable $d$ first.  Figure \ref{fig:ex7} shows chunk $C_{-1}$ and the inversion of $d$.  

\begin{figure}[tpb]
\centering
\input{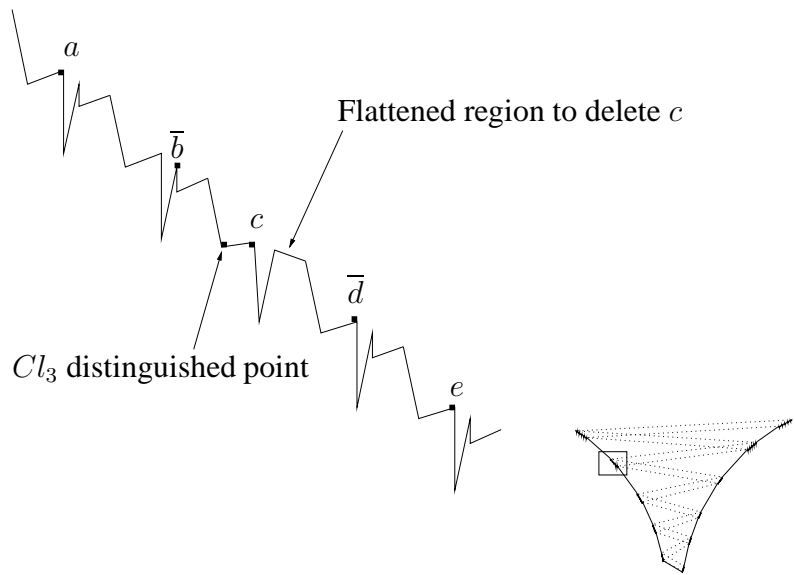}
\caption{Chunk $C_{-2}$ which places a clause gadget for clause $Cl_3$ and also deletes variable $c$.}
\label{fig:ex8}
\end{figure}

We can now place a clause gadget for $Cl_3 = (b \vee c \vee d)$.  We place this gadget in chunk $C_{-2}$ as shown in Figure \ref{fig:ex8}.  The mirrored distinguished point for $c$ is replaced with $Cl_3$'s distinguished point.  Since we are deleting $c$ in this chunk, we flatten out the other mirrored distinguished point for $c$.  We move $Cl_3$'s distinguished point accordingly so that it sees the $d$ guard location in chunk $C_{-3}$.  

The changes for the clause gadget are continued in chunk $C_{-3}$ as shown in Figure \ref{fig:ex9}.  The change we must make is adjusting the $m$ point as seen before.  The original purpose of $m$ was to ensure $b \in C_{-3}$ did not see other variables mirrored distinguished points to the right of $b \in C_{-2}$.  However, this changes because we want $b \in C_{-3}$ to see $Cl_3$'s distinguished point.  The only guard locations that see $Cl_3$'s distinguished point are $c \in C_{-2}$, $b \in C_{-3}$ and $d \in C_{-3}$.  We also delete the $e$ variable in chunk $C_{-3}$ to save space.  To delete $e$, the mirrored distinguished points for $e \in C_{-3}$ are flattened out.

\begin{figure}[tpb]
\centering
\input{bigExample9.pstex_t}
\caption{Chunk $C_{-3}$ which places a deletion gadget for variable $e$.}
\label{fig:ex9}
\end{figure}

\begin{figure}[tpb]
\centering
\input{bigExample10.pstex_t}
\caption{Chunk $C_{-4}$ which places a clause gadget for clause $Cl_4$ and also deletes variable $b$.}
\label{fig:ex10}
\end{figure}

\begin{figure}[tpb]
\centering
\input{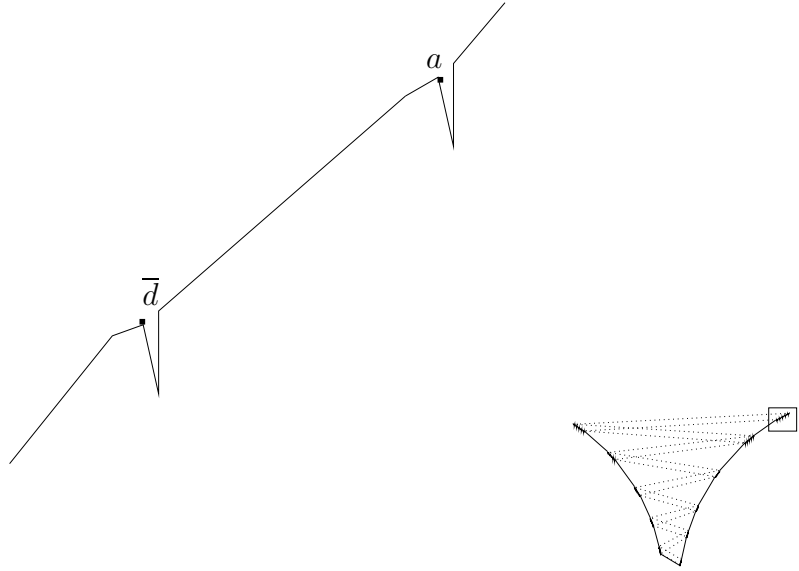}
\caption{Chunk $C_{-5}$ which places a deletion gadget for variables $a$ and $d$.}
\label{fig:ex11}
\end{figure}

Chunk $C_{-4}$ is shown in Figure \ref{fig:ex10}.  In this chunk we place the clause gadget for $Cl_4$.  We replace variable gadget $b \in C_{-4}$ with a clause gadget for $Cl_4$.  $Cl_4$'s distinguished point is adjusted accordingly so that it sees the $d$ literal guard location in chunk $C_{-5}$.  Since this is the last clause to be placed and to save space, the $a$ and $d$ variables are deleted in chunk $C_{-5}$ as shown in Figure \ref{fig:ex11}.  Adjustments to the terrain in chunk $C_{-5}$ are made similarly as before so that $a$ sees $Cl_4$'s distinguished point.  The only guard locations that see $Cl_4$'s distinguished point are $b \in C_{-4}$, $a \in C_{-5}$ and $d \in C_{-5}$.

This ends our reduction going upward and we count how many guards are necessary going upward.  We note that each variable gadget requires a guard to be placed at one of the literal points for that particular variable gadget because of the {\em Uniqueness Claim}.  No other point on the terrain sees these variable distinguished points so a guard is required to be placed at a literal location for each variable gadget.  We also add 1 extra guard for each inversion gadget that was placed.  Recall that an inversion requires 1 extra guard.  We count the number of variable gadgets in each chunk $C_{-1}, C_{-2}, C_{-3}, C_{-4},$ and $C_{-5}$ and end up with $(5+1)+5+4+3+2 = 20$.  

The entire terrain needs at least $39$ guards.  However $39$ guards are sufficient if the planar 3SAT instance is satisfiable.  Assuming correct choices were made in mirroring, no extra guards are required to see the mirrored distinguished points.  If the planar 3SAT instance is satisfiable, the entire terrain can be guarded with $39$ guards because the clause distinguished points will also be seen.  If more than $39$ guards are required to see the entire terrain, the planar 3SAT instance is not satisfiable.

\end{document}